\documentclass[]{svmult}

\usepackage{graphicx}        
\usepackage[bottom]{footmisc}

\begin{document}

\title*{Smart Focal Plane Technologies for VLT Instruments}
\titlerunning{Smart Focal Plane Technologies} 
\author{C. R. Cunningham \and C. J. Evans}
\institute{UK Astronomy Technology Centre, Royal Observatory Edinburgh, Blackford Hill, Edinburgh, EH9 3HJ, UK\\
\texttt{crc@roe.ac.uk; cje@roe.ac.uk}}
%
%
\maketitle

\section{Introduction}
\label{intro}

As we move towards the era of ELTs, it is timely to think about the
future role of the 8-m class telescopes.  Under the OPTICON programme
novel technologies have been developed that are intended for use in
multi-object and integral-field spectrographs.  To date, these have
been targeted at instrument concepts for the European ELT, but there
are also significant possibilities for their inclusion in new VLT
instruments, ensuring the continued success and productivity of these
unique telescopes.

\section{`Smart Focal Planes'}

A smart focal plane is a system that maximises the use of the
telescope focal plane for science observations.  Examples range from
wide-field, multi-object spectrometers and multiple
integral field units (IFUs), to future systems such as
robotically-fed, photonic spectrometers.  Under the Framework Six
OPTICON programme\footnote
{OPTICON has received research funding from the European Community's Sixth Framework Programme under contract number RII3-CT-001566.}, a wide range of novel systems has been developed by a team
from eight European countries \cite{crc}, providing a toolkit for instrument
designers including:
\begin{itemize}
\item{Replicated image slicers to enable more economic production of multiple
IFUs \cite{js}.}
\item{`Starbugs' -- miniature robots that carry fibres or pick-off mirrors to any
given place in the focal plane, potentially at cryogenic temperatures \cite{bugs}.}
\item{`Starpicker' -- an alternative cryogenic, robotic positioner to place pick-off 
mirrors on a potentially curved, dual focal-plane that is tumbled into
its observing position \cite{prh}.}
\item{Active correction mirrors used to correct for aberrations caused by the
movement of a pick-off mirror across the focal plane \cite{bsm}.}
\item{Micro-mirror arrays fabricated in silicon that can be used to form multi-object
slitlets with high densities of objects \cite{mma}.}
\end{itemize}

\section{Wide-field options for the VLTs}

In the era of ELTs, an obvious role for an 8-m telescope could be for
wide-field imaging surveys, in the same manner that CFHT and UKIRT
have been very productive in the 8-m era.  However, the VLTs were
never designed for wide-field operation and there are severe
optomechanical and operational constraints on providing a wide-field
facility at any of the VLT focal stations.  It will be very difficult
and expensive to extend the VLT focal plane to beyond one degree
\cite{spt}.  Indeed, wide-field imaging in the near-IR is not
competitive unless a field of at least one degree can be accessed; at
visible wavelengths even this would not be competitive with Pan-STARRS
and LSST.

A number of presentations at this workshop have made a clear case for
spectroscopy of a large number of sources over a field of a degree of
more, e.g. unprecedented galaxy redshift surveys to map large-scale
structure, or wide-area stellar surveys to trace the mass-assembly
history of the Milky Way.  Smart focal plane technologies can now deal
with the physical size, and possible curvature, of such a
field-of-view, even if cooled to cryogenic temperatures.  However, the
technical and cost problems relating to modification of the telescope
itself are severe.  The classical option for wide-field astronomy is
at prime focus, but the VLTs have significant mass, space and systems
limitations.  For instance, the secondary mirrors are part of the
active telescope system, and carry out real time tip-tilt and focus
correction at up to 50Hz. Any prime focus system would need to provide
similar functionality.

With the notable exception of Subaru, decisions taken in the early
design stages of the 8-m class telescopes have made prime focus
options difficult.  In principle, Gemini has the option for a
replaceable top-end to allow conversion to a wide-field mode, but this
has never been implemented; replacing the top end with a carbon-fibre
structure has been proposed to allow an increased mass at the prime
focus \cite{miz}, but using such a solution on one of the VLTs would
be very expensive.  It would also be necessary to provide correcting
optics -- a non-trivial challenge for an 8-m telescope.  For instance,
the Subaru corrector for the 30$'$ FMOS field (optimized from 0.9 to
1.8 $\mu$m) requires three elements of up to 600~mm diameter in BSM51Y
glass \cite{fmos}. Other options, such as the forward-Cassegrain
arrangement used to convert UKIRT to a wide-field mode for WFCAM
\cite{wfcam}, would likely be even more challenging for the VLT.

\section{Cladistics \& natural selection}

\begin{figure}
\centering
\includegraphics[viewport=1 100 600 750,height=13cm]{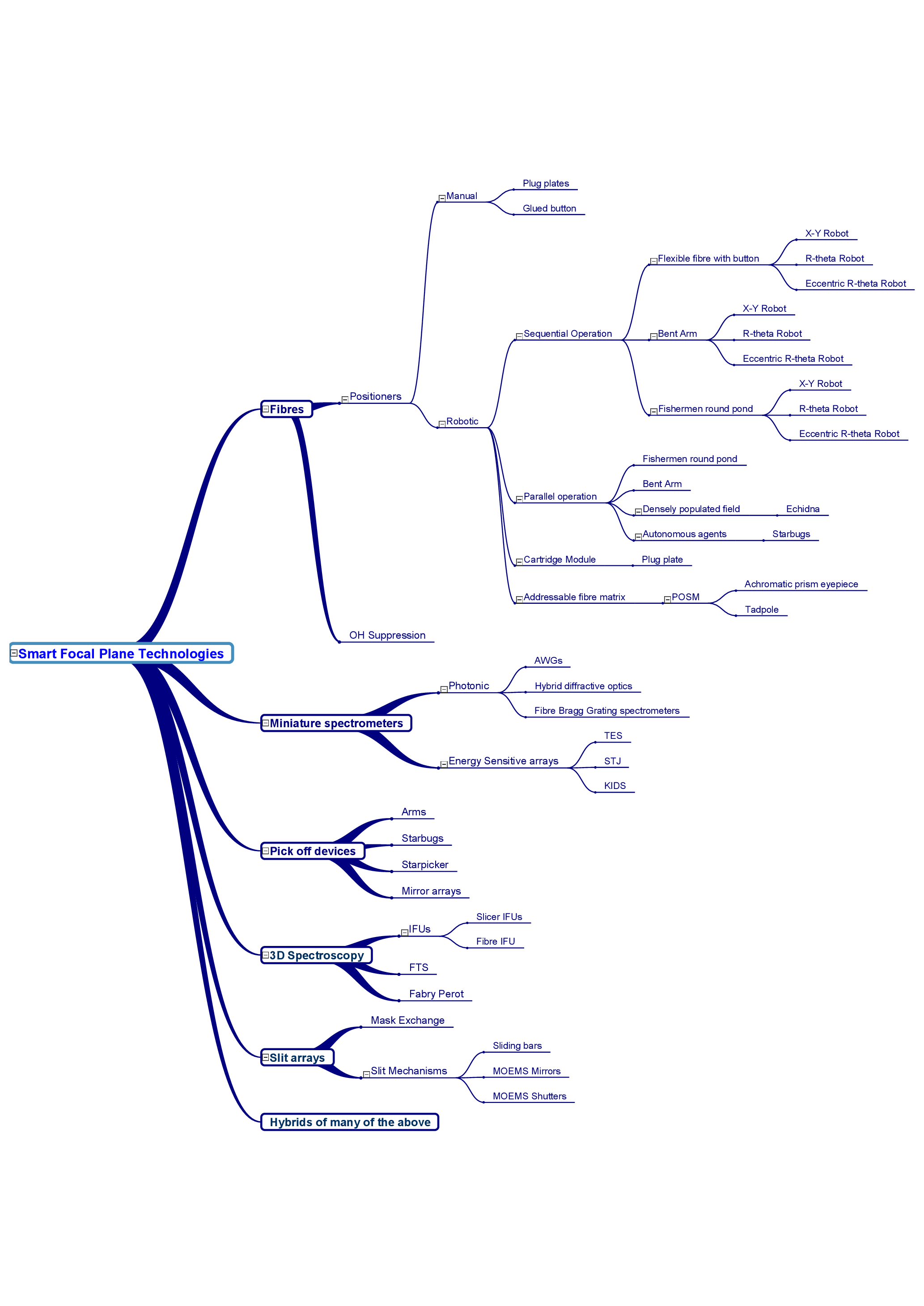}
\caption{Cladistic map for multi-object spectroscopy.  Adapted from \cite{smith}.}
\label{fig1}
\end{figure}

It can be argued that technology selection for astronomical
instruments follows a similar process to natural selection in nature.
Selection pressures have analogies in terms of resources, competition
and even predation -- although it is hard to see where sexual
selection takes a role!  Many ideas follow an evolutionary path with
ideas splitting-off from an original concept, through an almost
cladistic mapping, as illustrated by the example of wide-field spectroscopy
(Figure~\ref{fig1}), where most systems can be traced back to Medusa,
the first fibre plug-plate system at Steward Observatory \cite{hill}.

It is noteworthy how evolution has led to significant novelty in an
isolated continent like Australia.  Similarly, the team at the
Anglo-Australian Observatory (AAO) have been responsible for many
innovations!  A recent example is the Echidna positioner developed for
FMOS, that deals with dense packing of targets within a physically
small field, in which traditional pick-and-place devices are not
feasible \cite{fmos}.  A similar design to the Echidna positioner has
been proposed for Gemini/Subaru-WFMOS, while an alternative concept
uses small field lenses to select sub-fields which are then fed by
miniature pick-offs to a fixed grid of fibres.  A further concept
proposed at this workshop is an update of the traditional fibre
plug-plate, using robotics to assemble replaceable fibre modules \cite{parry}

\section{Multi-slit spectroscopy}

The science case for slit spectroscopy remains strong, particularly in
cases where both a high multiplex and good throughput are required for
point-like sources with known positions.  New technologies are now
becoming available to replace slit masks with configurable
slit-arrays, that can also operate at cryogenic temperatures.  The
slitlet mechanism developed for the European Space Agency by CSEM 
under the OPTICON smart focal planes programme is now being
incorporated into the Keck MOSFIRE instrument \cite{mosfire}.
Moreover, devices such as micro-mirror arrays \cite{mma} and the
shutter arrays in the {\it JWST} NIRSPEC instrument \cite{nirspec}
offer potential arrays of up to 10,000 slitlets.  These could be used
in concepts such as the proposed VLT-MegaMOS \cite{mega}.

\section{IFU spectroscopy}

In many cases, such as attempts to disentangle observations of merging
galaxies, we require three-dimensional imaging spectroscopy.  In this
instance, image slicers offer significant advantages over slits.
Precision manufacturing techniques such as diamond machining,
stacked-element glass slicers and electroformed replication could
provide a larger number of channels, or larger field-of-view, than the
capabilities of KMOS or MUSE. When combined with new OPTICON pick-off
technologies (e.g. Starbugs and Starpicker) instruments can be forseen
that combine a relatively wide patrol field, with high spatial
resolution (i.e. fed by future AO systems).

\section{Miniature spectrometers}

In the longer-term, new technologies from the photonics industry could
result in integrated, miniature spectrometers.  For instance, array
waveguide devices could be used if the problems of efficient
light-coupling and bandwidth can be solved \cite{horton}.  There are
also integrated devices under development for fluorescence
spectroscopy that are based on bulk optics and holographic gratings,
yielding spectral resolutions of up to 400 at visible wavelengths from
a device that can fit within a 1\,cm cube \cite{mini}.  These concepts
may be useful in a highly-multiplexed instrument, in which a swarm of
miniature spectrometers are self-propelled around the focal plane by
autonomous robotic carriers.

The ultimate miniature imaging-spectrometer would be an energy
sensitive detector. Unfortunately, current devices such as
superconducting tunnel junction arrays, or transition edge
superconductors, do not offer large pixel-densities, nor useful
spectral resolution.  However, it is possible that such
components could revolutionise spectrometer design in the future, possibly
in conjunction with photonic feed devices.

\section{Conclusions}
In summary, there are considerable technical challenges in developing
a wide-field spectrograph for the VLT, particularly when one considers
the available resources within ESO and of its partners; a wide-field
prime focus instrument would also lead to operational problems for
VLTI.  Future smart focal plane and photonics technologies may enable
a lighter, more compact, prime focus (or foward-Cassegrain)
instrument, but current technology readiness levels suggest that a
compromise would be to exploit the maximum field available at the
Nasmyth foci, e.g. Super-Giraffe \cite{giraffe}.  Meanwhile,
technology development for ELT instrument concepts such as EAGLE, and
through the next Framework Programme, should be supported to help
break the existing paradigm.

\end{document}